\documentclass{aa} 
\usepackage{epsfig,times} 

\def\mincir{\raise -2.truept\hbox{\rlap{\hbox{$\sim$}}\raise5.truept \hbox{$<$}\ }} 
\def\mincireq {\hbox{\raise0.5ex\hbox{$<\lower1.06ex\hbox{$\kern-1.07em{\sim}$}$}}} 
\def\magcir{\raise -2.truept\hbox{\rlap{\hbox{$\sim$}}\raise5.truept \hbox{$>$}\ }} 
\def\gr{\kern 2pt\hbox{}^\circ{\kern -2pt K}} 
\def\e{\epsilon}

\def\IRAS{{\sl IRAS~}}

\def\a{\alpha}

\def\e{\epsilon}

\def\G{\Gamma}

\def\W{\Omega}
\def\L{\Lambda}
\def\t{\tau}
\def\s{\sigma}

\def\CXB{\rm CXB}

\def\TB{\rm TB}

\def\cPL{\rm cPL}

\def\_{\thinspace}

\def\keV{\thinspace{\rm keV}}

\def\Mpc{\thinspace{\rm Mpc}}

\def\sec{\thinspace{\rm s}}
\def\sr{\thinspace{\rm sr}}

\def\cm{\thinspace{\rm cm}}
\def\cm2{\thinspace{\rm cm}^2}

\def\be{\begin{equation}}
\def\ee{\end{equation}}

\begin{document} 
 
\title{Starburst galaxies and the X-ray background}  
 
\author{Massimo Persic\inst{1} and 
	Yoel Rephaeli\inst{2,3}} 
 
\offprints{M.P.; e-mail: {\tt persic@ts.astro.it}} 
 
\institute{ 
INAF/Osservatorio Astronomico di Trieste, via G.B.Tiepolo 11, 34131 
Trieste, Italy
	\and  
School of Physics and Astronomy, Tel Aviv University, Tel Aviv 69978, 
Israel
	\and
CASS, University of California, San Diego, La Jolla, CA 92093, USA}
\date{Received ..................; accepted ...................}

\abstract{Integrated X-ray spectra of an evolving population of
starburst galaxies (SBGs) are determined based on the observed
spectra of local SBGs. In addition to emission from hot gas and 
binary systems, our model SBG spectrum includes a nonthermal 
component from Compton scattering of relativistic electrons by the 
intense ambient far-IR and the (steeply evolving) CMB radiation 
fields. We use these integrated spectra to calculate the levels of 
contribution of SBGs to the cosmic X-ray background assuming that 
their density evolves as $(1+z)^q$ up to a maximal redshift of 5. 
We find that at energies $\e \mincir 10$ keV this contribution is 
at a level of few percent for $q \leq 3$, and in the range of 
$5\%-15\%$ for $q \simeq 4.5$. The Compton component is predicted 
to be the main SBG emission at high energies, and its relative 
contribution gets progressively higher for increasing redshift.   
}

\maketitle 
\markboth{Persic \& Rephaeli: SBG contribution to the CXB}{}

\section{Introduction} 

In starburst galaxies (SBGs) enhanced star formation activity
(lasting typically $\mincir 10^8$ yr) drives a chain of coupled
stellar and interstellar phenomena that are manifested
by intense far-infrared (FIR) emission. The SBG spectrum shows
a distinctive large bump of thermal FIR dust emission which
is correlated with optical emission (e.g., Silva et al. 1998), 
a result of the fact that
dust emission is reprocessed starlight from a population of
hot OB stars. Such stars are short-lived ($10^{6-7}$ years) and
end up as supernovae. The subsequent radio emission from synchrotron 
radiation in supernova remnants (SNRs) is tightly correlated with 
the IR emission (e.g., Condon 1992). Interest in SBGs stems also from
the realization that these resemble young galaxies in the earlier
universe. Indeed, a starburst phase was very common then, as a result of
both astrophysical processes (baryonic infall and early star
formation) and dynamical processes (close encounters and mergers
which, in turn, trigger star formation). Consequently, the cosmic
star formation rate (SFR) was substantially higher at epochs
corresponding to $z \magcir 1$, with the SFR having either a peak
at $1 \mincir z \mincir 2$ (Madau et al. 1996), or a plateau out to
$z \sim 4$ (Thompson et al. 2001).

A primary manifestation of the starburst activity is X-ray emission. Given
the greatly enhanced SFR, energetic phenomena related to stellar
evolution -- OB stars, X-ray binaries, SNRs, galactic winds, and
Compton scattering of ambient FIR and CMB photons off relativistic
electrons that are accelerated by SN shocks -- clearly suggest that
SBGs are typically more powerful X-ray emitters than normal galaxies
of comparable stellar mass (e.g., Rephaeli et al. 1991; Schmitt et al. 
1997). The mean X-ray spectrum of SBGs reflects the diverse nature of 
high energy activity in these galaxies (Persic \& Rephaeli 2002; 
Rephaeli et al. 1991, 1995).

The link between SBGs and the cosmic X-ray background (CXB) was made
early on (Bookbinder et al. 1980; Stewart et al. 1982; Weedman 
1987; Griffiths \& Padovani 1990; Rephaeli et al. 1991, 1995; 
David et al. 1982; Ricker \& Meszaros 1993; Moran et al. 1999). 
Most previous estimates of the SBG contribution to the CXB were 
essentially based on the SBG emission at 2 keV, with either no source 
evolution (Weedman 1987), or with some assumed degree of evolution 
(Griffiths \& Padovani 1990; Ricker \& Meszaros 1993). Clearly, the 
use of the 2 keV emission as baseline is unjustified given the wide 
spectral range (3-100 keV) over which the CXB seems to have been well 
determined (e.g., Gruber et al. 1999). The SBG contribution was also 
evaluated based on the emission over a wider spectral range, either 
in the context of an assumed spectral model (Ricker \& Meszaros 1993), 
or from a statistically deduced mean spectrum (determined from the 
{\sl HEAO-1} A2+A4 datasets: Rephaeli et al. 1991, 1995).

In this paper we compute the contribution of SBGs to the CXB incorporating 
recent results in the study of SBGs. In particular, we 
use the knowledge of SBG spectral properties gained from direct {\sl 
ASCA}, {\sl BeppoSAX}, and {\sl RXTE} observations of local galaxies.
This, complemented with an assumed form for the cosmic evolution of the
SBG X-ray luminosity density, enables us to compute the component of 
the CXB spectrum which originates in SBGs. We first (section 2) 
review the spectral properties of (local) SBGs, and proceed
to discuss the local SBG luminosity density and its cosmic evolution 
in sections 3 and 4. The predicted CXB spectrum is then calculated 
in section 5, followed by a discussion (section 6), and a summary 
(section 7). Our calculations are carried out in the context of the 
Einstein-de Sitter ($\W_m=1$, $\W_\L=0$; hereafter EdS) and  
(currently favoured) flat $\Lambda$ ($\W_m=0.3$, $\W_\L=0.7$) cosmological 
models, with $H_0=50$ km s$^{-1}$ Mpc$^{-1}$.

\section{X-ray spectral properties of SBGs} 

Local SBGs have been repeatedly observed in X-rays (see Fabbiano 
1989; Rephaeli et al. 1995; Dahlem et al. 1998; Persic \& Rephaeli 
2002; and references therein). In
most of these ($\sim 10$) galaxies the 0.5-10 keV spectra can be well
fit by a combination of thermal, low temperature $kT \simeq 0.8$ keV
emission, and photoelectrically self-absorbed emission that can be
represented by an exponentially cutoff power law (CPL) of the form
$x^{1-\G} e^{-x/kT}$, where $x$ is the energy in the source frame, 
$\G$ is the photon index and $kT$ is the cutoff energy 
         \footnote{
         Earlier spectral studies, in which simple thermal and 
	 PL models were fitted to the main emission component, 
	 concluded that $\magcir$5 keV thermal or $\G \sim 2$ 
	 PL fits were similarly successful for all analyzed 
	 objects (NGC253, M82: Ptak et al. 1997, Cappi et al. 1999;
         M83: Okada et al. 1997; NGC1569: Della Ceca et al. 1996; 
	 NGC2146: Della Ceca et al. 1999; NGC2903: Mizuno et al.
	 1998; NGC3256: Moran et al. 1999; NGC3310, NGC3690: Zezas 
	 et al. 1998), and hence they were unable to reach definitive 
	 conclusions on the nature of the main component.}
(Persic \& Rephaeli 2002, Persic et al. 2002; see Fig.1).

The low-temperature and CPL components can be interpreted as due to
emission from galactic winds and X-ray binaries, respectively. 
From a detailed discussion of a synthetic X-ray spectrum of SBGs, 
based on an evolutionary model of galactic stellar populations and on 
the X-ray spectra of the relevant emission processes, Persic \& 
Rephaeli (2002) have suggested that X-ray binaries contribute most 
of the 2-15 keV emission. Both types of X-ray binaries, high- and 
low-mass systems, have spectra that can be described as variously 
cutoff power laws (White et al. 1983; Christian \& Swank 1997). In 
general, the population-averaged spectrum emitted by a realistic 
mix of X-ray binaries (high- and low-mass systems of various
luminosities in Galactic proportions) can be described as a CPL 
with $\G \sim 1.2$ and $kT \sim 8$ keV (Persic \& Rephaeli 2002).

The enhanced SFR activity in SBGs results in a high SN rate and 
consequently -- among other effects -- in a higher rate of electron 
acceleration to relativistic energies. This prospect, coupled with 
the higher energy density in the enhanced FIR radiation field, 
almost certainly implies that Compton scattering of the electrons 
by both the FIR and CMB fields yields a substantially higher level 
of nonthermal X-ray emission than in normal galaxies (Schaaf et al. 
1989; Rephaeli et al. 1991). A quantitative assessment of this
process in NGC253 was given by Goldshmidt \& Rephaeli (1995).
Substantial PL X-ray emission may also be produced in the galactic 
nucleus, especially if the SB-driven turbulence of the gas increases 
the mass inflow rate onto a central black hole (e.g., Veilleux 2001). 
Such spectral components can be described as $\propto x^{1-\a}$ with 
the photon index in the range $\a \sim 1.6-1.8$ over a broad energy 
range (e.g., Goldshmidt \& Rephaeli 1995; Rothschild et al. 1983).

The spectral similarity shown by the local SBGs (see also Roberts 
et al. 2001) provides a reasonably good basis for a first 
approximation at modelling the population as a whole. We therefore 
{\it assume} that a typical starburst spectrum consists of a 0.8 
keV thermal component [i.e., a bremsstrahlung spectrum, with the 
Gaunt factor numerically calculated as prescribed by Itoh et al. 
(2000)], plus a CPL with photon spectral index $\G=1.2$ and cutoff 
energy $kT=7.5$ keV. This second component is photoelectrically 
absorbed {\it in situ} through a HI gas of column density $N_{\rm H}$. 
The two components are normalized such that the thermal-to-CPL 
energy flux ratio in the 2-10 keV band is 0.03. 

In addition to the wind and binary emission, we include also two 
emission components originating from Compton scattering of 
relativistic electrons by the FIR and CMB radiation fields. While 
the full calculation of the respective fluxes  requires a detailed, 
self-consistent solution of the kinetic equation for the electrons 
(taking into account their varius energy loss mechanisms and their 
propagation mode in the disk and halo -- see, e.g., Rephaeli 1979), 
we greatly simplify the treatment by keeping only the most salient 
features pertinent to our discussion here. Assuming the typical 
value of $0.8$ for the radio (spectral energy) index of the radio 
synchrotron flux in the disk, and estimating typical energy loss 
times of the electrons by synchrotron emission and Compton scattering 
off the FIR and CMB radiation fields, we can roughly represent the 
latter two Compton components as follows: A primarily inner-disk 
FIR-scattered component with a (photon) flux that is $\propto 
x^{-1.8}$, and a (primarily) outer-disk and halo CMB-scattered 
component $\propto x^{-2.3}$ due to electrons from the higher-energy 
region of the (electron) spectrum that is steepened by radiative losses. 

The relative contribution of the Compton fluxes is substantially 
uncertain since this emission has not yet been unequivocally detected 
in SBGs (see, e.g., Goldshmidt \& Rephaeli 1995 and Rephaeli \& Gruber 
2002). However, the expectation that Compton scattering of the 
enhanced relativistic electron population by the intense ambient FIR 
radiation field provides strong motivation for inclusion of this 
emission in our modelling of SBG X-ray emission. Our modelling of the 
Compton fluxes is based on the radio properties of the nearby SBGs 
M82 and NGC253, and well known Compton-synchrotron relations (e.g., 
Tucker 1975) between the radio luminosity and deduced mean magnetic 
field in the disk (typically, $\sim 10$ $\mu$G, but with appreciable 
uncertainty) and the predicted X-ray flux from Compton scattering of 
the radio producing electrons by the FIR and CMB fields. The relative 
strength of these two radiation fields largely determines (under 
typical conditions in local SBGs) the normalization between their 
respective fluxes. The energy density of the FIR radiation field, 
$U_{\rm fir}$, was estimated by Goldshmidt \& Rephaeli (1995) in the 
context of a quantitative model for the emission from warm dust. We 
adopt their result, $U_{\rm fir} \simeq 10 \, U_{\rm 0}$ where $U_{\rm 
0}$ is the energy density in the CMB, as roughly typical in the inner 
region of a SBG disk. Doing so, and using the usual Compton-synchrotron 
relations, we obtain that in local SBGs we might expect that the FIR 
and CMB Compton contributions to the overall 2-10 keV (energy) 
flux are roughly at a level of $5\%$ and $0.5\%$, respectively. 
We do not yet have unequivocal evidence for nonthermal Compton X-ray
emission in SBGs. Power-law fits to both ASCA and RXTE data on M82 \& NGC253
were found to be formally acceptable (e.g.: Matsumoto \& Tsuru 1999; 
Rephaeli \& Gruber 2002), but a closer assessment of the residuals of such 
fits over the wider energy range afforded by RXTE leads only to what are
essentially upper limits to nonthermal contribution. With the higher
spatial resolution of the {\sl Chandra} satellite it was clearly
established that some of the 2-10 keV emission in M82 is diffuse,
emanating from a central elliptical region, 0.8 kpc $\times$ 0.8 kpc,
that is coplanar and coaligned with the galactic disk (Griffith
et al. 2000). According to these authors the spectral fits to these
measurements make a Compton origin of most of the emission quite likely.
The luminosity of the extended emission region is $L_{2-10} \simeq
2.2\times 10^{39}$ erg s$^{-1}$ (for an assumed distance $D=3.6$ Mpc);
this value constitutes $\sim 5\%$ of the total 2-10 keV luminosity
($L_{2-10} \simeq 4.5 \times 10^{40}$ erg s$^{-1}$) measured by
{\sl BeppoSAX} (Persic \& Rephaeli 2002). 
Given that some or most of this diffuse hard component in M82 is thermal 
as suggested by the presence of a substantial 6.7 keV Fe-K emission, the 
figure of $5\%$ is actually a strong upper limit on the Compton contribution 
in M82.
	\footnote{Also for NGC 3256 (the brightest among local SBGs),
	Compton scattering has been argued to contribute appreciably 
	to the 2-10 keV flux (Moran et al. 1999).}

Finally, let $f(x)$ be the spectral profile (in the source frame), 
normalized in the band relative to which the surface brightness is 
being computed. In accord with our discussion above, the X-ray spectral 
profile of a SBG is a superposition of components that can be written 
as 
$$
f(x) ~=~ 
\beta \, f_{\TB}(x) ~+~ 
(1-\beta) \, e^{-N_{\rm H} \sigma_{pa}(x)} f_{\cPL}(x) ~+~  
$$
$$
~+~ \eta_1 \, e^{-N_{\rm H} \sigma_{pa}(x)} f_{\rm C}^{\rm FIR}(x) ~+~  
\eta_2 \, f_{\rm C}^{\rm CMB}(x,z)
\eqno(1)
$$
where $x \equiv \e \,(1+z)$ is the energy (in keV) in the source
frame (and $z$ is the redshift; $\e$ is the energy in the observer's 
frame), $\sigma_{pa}$ is the cross section for photoelectric 
absorption per unit hydrogen column density $N_{\rm H}$ within the source 
[in the calculations we have adopted the analytic expression for 
$\sigma_{pa}$ given by Morrison \& McCammon (1983); $N_{\rm H}$ is taken 
at a nominal level of $10^{22}$ cm$^{-2}$], and $\beta$, $\eta_1$, 
and $\eta_2$ are constants of normalization that are estimated as 
described above. Note that the CMB scattered flux is $z$-dependent, 
an important fact that will be elaborated upon in Section 5.

\begin{figure}
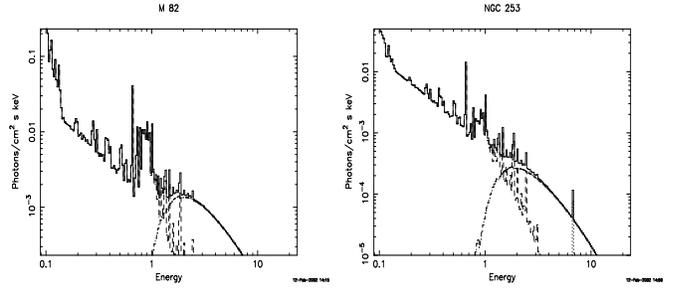

\vspace{1.8cm}
\includegraphics{papAA3067_fig1a.ps}
\hskip 0.2cm
\vspace{1.8cm}
\includegraphics{papAA3067_fig1b.ps}
\caption{
The best-fit thermal + CPL models for M82 (left) and NGC253 (right), 
based on {\it BeppoSAX} data (see Persic \& Rephaeli 2002).}
\end{figure}

\begin{figure}
\vspace{4.0cm}
\includegraphics{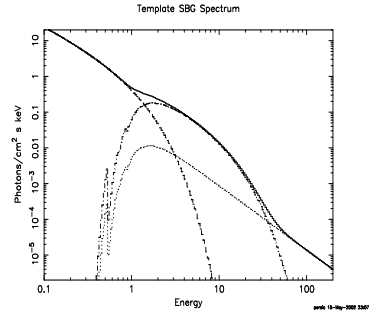}
\caption{
The template photon SBG spectrum assumed in the present calculations.
It consists of:
{\it (i)} an unabsorbed 0.8 keV thermal bremsstrahlung component, 
{\it (ii)} an exponentially cutoff PL with photon index $\G=1.2$ and
cutoff energy $kT=7.5$ keV, absorbed through $N_{\rm H}=10^{22}$ cm$^{-2}$,
and {\it (iii)} a similarly absorbed PL with photon index 
$\alpha_1=1.8$. (The much fainter unabsorbed PL with photon index 
$\alpha_2=2.3$ is not shown here.) The flux normalizations are as 
specified in the text.}
\end{figure}

\section{Local luminosity density} 

In order to calculate the contribution of SBGs to the CXB we
need to model the population at all $z$, beginning -- of course --
with the local source density which can be deduced from the FIR
luminosity function. This has been determined from the 60$\mu$m
flux-limited \IRAS samples (Saunders et al. 1990),
$$
\phi(L) \, {d\, {\rm log}L}   ~=~
C\, \biggl({L \over L^\star}\biggr)^{1-\alpha} e^{-{1 \over 2\, \s^2}
{\rm log}^2(1+L/L^\star) } {d\,{\rm log}L} \,.
\eqno(2a)
$$
Sources in the \IRAS Point Source Catalog with fluxes measured in all
four \IRAS bands and with individual, identifiable galaxy types inhabit
distinct and well-defined areas in color-color diagrams (Rowan-Robinson
\& Crawford 1989). Each one of these distinct components of the \IRAS 
point-source population largely produces most of the emission at 
specific wavelengths, and is characterized by a specific set of 
parameters for eq.(2$a$). For the warm-source component that peaks 
at $\sim$60$\mu$m which is thought to be attributable to a dusty SBG 
population, the parameters in eq.(2$a$) take the values:
$$
\left\{
\begin{array}{ll}
C= 3.25 \times 10^{-4} \, {\rm Mpc}^{-3} & \mbox{~~~~~~~~~~~~~~} \\
{\rm log}\,L^\star=9.99  & \mbox{~~~~~~~~~~~~~~} \\
\alpha= 1.27 & \mbox{~~~~~~~~~~~~~~} \\
\s = 0.626 & \mbox{~~~~~~~~~~~~~~} \\
\end{array}
\right.
\eqno(2b)
$$
for luminosities in the range $6 \leq {\rm log}(L_{60}/L_\odot) \leq
12.5$. Pearson \& Rowan-Robinson (1996) modeled the (sub-mm, IR,
optical) number-flux relation for extragalactic sources by including
a SBG population described by a local luminosity function as in eq.(2).
The local 60$\mu$m luminosity density implied by eq.(2) is:
$$
{\cal L}_{60\mu{\rm m}} = 8.4 \times 10^6 L_\odot \Mpc^{-3}\,.
\eqno(3)
$$
For a sample of local SBGs observed with {\sl ASCA}, the ratio of
2-10 keV flux to 60$\mu$m flux
        \footnote{
        The objects are: M82, NGC253, NGC2903, NGC3690. The 60$\mu$m
        fluxes are calculated using the prescription $F(60)= 2.58
        \times 10^{-14} f_\nu(60)$, where $f_\nu$ is the nominal flux 
	densities listed in the \IRAS catalog (Lonsdale et al. 1985; 
	see also Shapley et al. 2001). The 2-10 keV fluxes are from 
	Ptak et al. 1997 (M82, NGC253), Mizuno et al. 1998 (NGC2903), 
	Zezas et al. 1998 (NGC3690).}
is $(4 \pm 1) \times 10^{-4}$. Assuming this to be the mean value for 
the whole SBG population, we have for the 2-10 keV luminosity density
$$
{\cal L}_{2-10\, {\rm keV}}(0) ~=~ (1.3 \pm 0.3) \times 10^{37} 
~~ {\rm erg ~s}^{-1} {\rm Mpc}^{-3} \,.
\eqno(4)
$$

Note that in our previous estimates of the SBG contribution to the CXB 
(Rephaeli et al. 1991, 1995) we did consider only the high end ($L_{\rm 
FIR} \geq 10^{11} \, L_\odot$) of the FIR luminosity function as 
representative of the SBG population. Adopting the current more inclusive 
luminosity function (Pearson \& Rowan-Robinson 1996) allows a more realistic 
estimate that accounts also for the emission from fainter SBGs.
By integrating the FIR luminosity function over the above (logarithmic) 
range ([6, 12.5]), we compute a density of $\simeq 5.9 \times 10^{-3}$ 
Mpc$^{-3}$ for galaxies that are identified here as members of the 
broader SBG population. (Obviously, this value of the density is much 
higher than that of the more luminous galaxies, but this difference is 
largely compensated for by the much lower mean FIR luminosity, $L_{\rm 
FIR} \sim 10^{9} \, L_\odot$, of the SBGs that we include in this work.)

\section{Cosmic evolution} 

Galactic evolution may generally be reflected in density and luminosity
variations over cosmic time due to the continued formation, interaction
and merging, and through intrinsic changes of the luminosity due to
dynamical and thermal evolution of stars and interstellar gas. The
evolution of SBGs is likely to be particularly pronounced as the SB
phase is thought to be triggered by galactic interactions and is
relatively short lived. These imply that SBGs were more abundant at
earlier times than would be expected based solely on pure density
evolution of sources in an expanding background ($\propto (1+z)^3$).

Observationally, the evolution of SBGs is inferred from radio, IR, and
optical measurements. At radio frequencies, two distinct populations are
identified at 1.4 GHz (Benn et al. 1993): radio-loud, and radio-faint,
the first consisting of giant ellipticals and QSOs, and the second 
mainly of spirals (SBGs and radio-quiet QSOs). This latter faint 
population is responsible for the low-flux upturn in the differential 
counts, and is indistinguishable from the SBG-dominated 60$\mu$m 
population (e.g., Danese et al. 1987; Franceschini et al. 1988), 
which can be fit by a model using pure luminosity evolution of the 
form $L(z)=L(0)\, (1+z)^q$ with $q=3.1$ (Rowan-Robinson et al. 1993). 
At IR wavelengths, Lonsdale et al. (1990) concluded that the dominant 
population in the faint 60$\mu$m counts from the \IRAS Faint Source 
Survey consists of strongly evolving SBGs.
        \footnote{In ultra-luminous IR galaxies (ULIRGs), which are the 
	high-$L_{60\mu{\rm m}}$ tail of the 'warm' 60$\mu$m-selected 
	\IRAS population, the star-formation activity (Kim \& Sanders 
	1998a) is very strong; these undergo cosmic evolution (Kim \& 
	Sanders 1998b) similar to the one derived for the whole 
	60$\mu$m-selected SBG sample. In addition, a preliminary {\sl 
	XMM-Newton} study of a sample of ULIRG spectra (Franceschini et 
	al. 2002) suggests properties fairly similar to those of SBG 
	spectra: i.e., the `starburst signature' discussed in section 2, 
	plus sometimes a hard PL component (interpretetd as SB-obscured 
	AGNs). ULIRGs are then implicitly included in the calculations 
	reported in this paper.}
Thus, it appears that from 60$\mu$m to radio wavelengths what is mainly 
sampled is starburst activity. 
SBG evolution, either of the pure luminosity type (Lilly et al. 1995; 
Ellis et al.1996; Fried et al. 2001), or of the pure density type (Lin et 
al. 1999), is deduced also from optical surveys.
\begin{figure}
\vspace{6.0cm}
\includegraphics{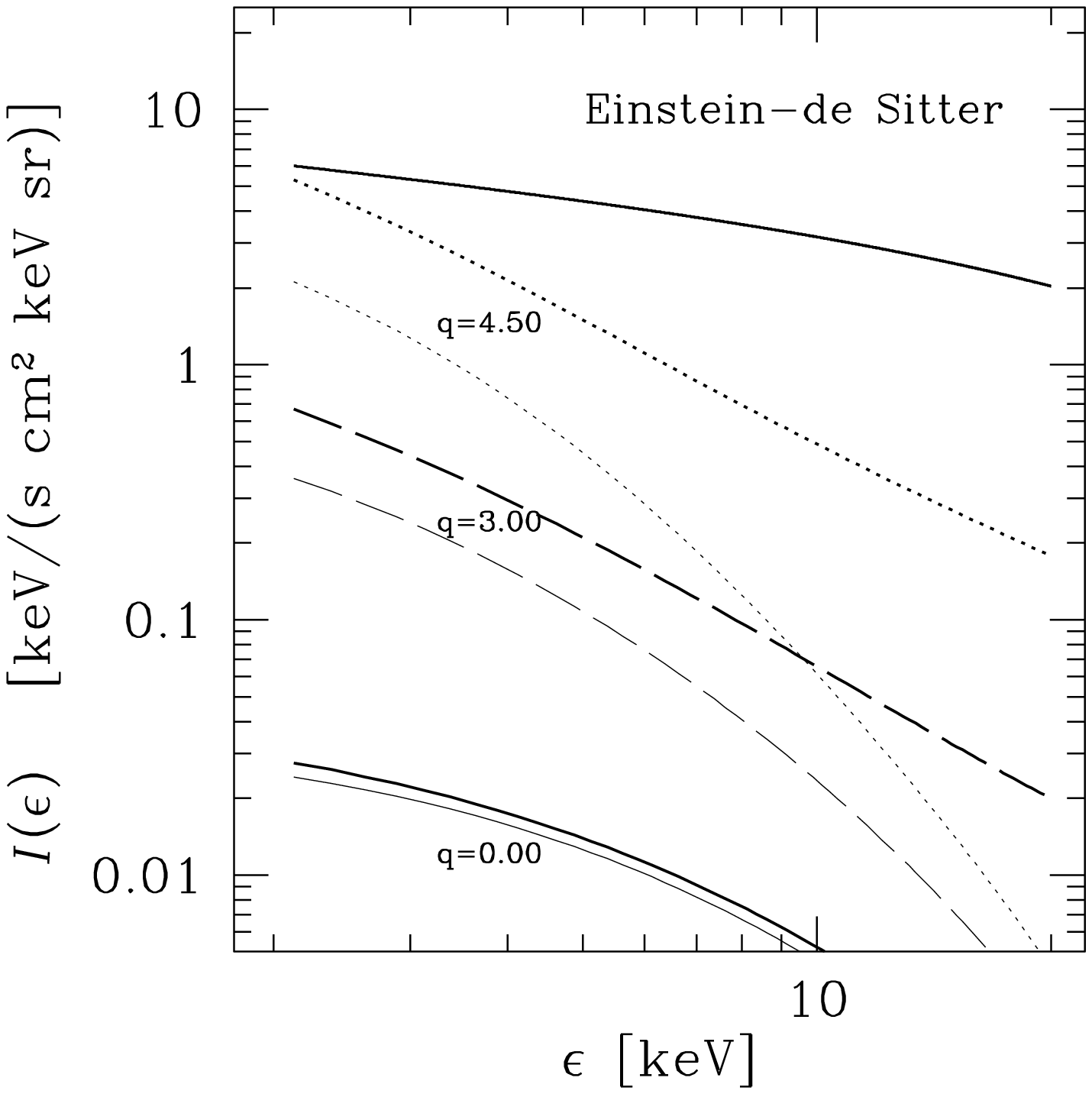}
\vskip 0.2cm
\vspace{6.0cm}
\includegraphics{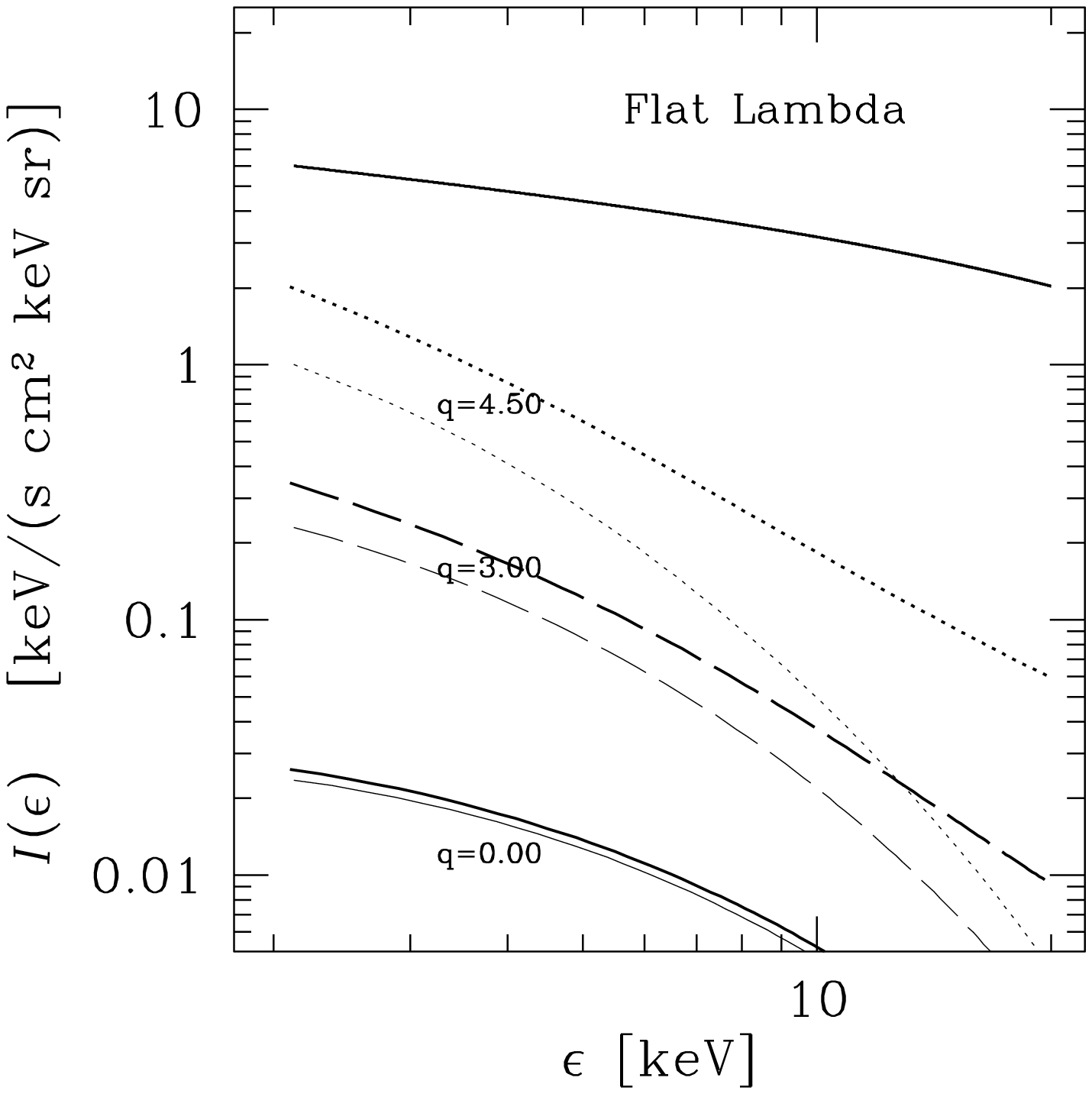}
\caption{
The spectrum of the integrated background light arising from SBGs
evolving out to $z_m=5.0$, for different values of the evolution
parameter. For each of the three values $q=0$, 3, 4.5, indicated by 
solid, long-dashed, and dotted curves, the lighter curve represents 
the unabsorbed bremsstrahlung emission plus absorbed CPL, 
and the heavy curve
shows the sum of these two components plus Comptonized 
emission (which consists of an absorbed $\a_1=1.8$ PL and unabsorbed 
$\a_2=2.3$ PL). 
The local SBG X-ray luminosity density is given in eq.(4), its cosmic
evolution is given in eq.(5), and the template SBG spectrum is shown
in Fig.2. The solid black curve on top represents the observed CXB 
[see eq.(11)].}
\end{figure}

The luminosity density of SBGs evolves mainly as result of their 
steeply varying density, and -- to a smaller extent, as discussed 
below -- the changes in the relative strength of the spectral 
components. Density evolution is represented by the factor $(1+z)^q$, 
with $q \geq 3$ after the process of galaxy formation began in ernest 
at some early redshift, $z_m$. Since we do not know when exactly 
galaxies began forming, the modelling the starburst phase of early galaxies 
necessitates viewing both $q$ and $z_m$ as essentially free parameters.
        \footnote{
        As an example, we derive an evolution function of the type in
        eq.(5) within a toy model of galaxy evolution. The comoving
        number density of galaxies undergoing a starbursting phase at any given
        epoch $z$, $n_{\rm SBG}(z)$, is closely related to the total
        galaxy density at that time, $n_{\rm G}(z)$. If so, we expect
        the comoving SBG fraction to increase steeply with $z$. The
        interaction rate between galaxies is $\propto n_{\rm G}(z)^2 v$
        (with constant interaction cross section), where $n_{\rm G}(z)$
        is the galaxy density, and $v \equiv \sqrt
        {\bar {(\delta v)^2}}$ is the rms pairwise velocity dispersion
        of interacting galaxies. Because of the strong $n_{\rm G}
        \propto (1+z)^3$ dependence we can ignore the relatively weak 
	dependence on $v$. This implies that in a flat universe the 
	volume density of SBGs at any given time is $n_{\rm SBG}(z) =
        n_{\rm SBG}(0)(1+z)^{4.5}$.}
We factor out the dominant redshift dependence of the luminosity 
density:
$$
{\cal L}(z) ~=~ {\cal L}(0) \, (1+z)^q \, \phi(z) \,,
\eqno(5)
$$
with all the spectral dependence on $z$ collected in the function 
$\phi(z)$; this function will be specified below. The evolution of SBGs 
is then largely gauged by the index $q$
        \footnote{
        Evidence for evolution of the galaxy luminosity density has been
        found also in other bands. Based on $z \leq 1.5$, {\it B}- and 
        {\it UV}-selected galaxies, the deduced evolution rate has 
        $q = 2.4 \pm 1.0$ for an EdS cosmology and $q=1.7 \pm 1.0$ for 
	a flat $\L$ model (Wilson et al. 2002). Earlier work, based on 
	$z\leq 1$, {\it I}-band selected galaxies, claimed $q \mincir 4$ 
	(Lilly et al. 1996).}; 
an estimate for ${\cal L}(0)$ is given in eq.(4).
\begin{figure}
\vspace{6.0cm}
\includegraphics{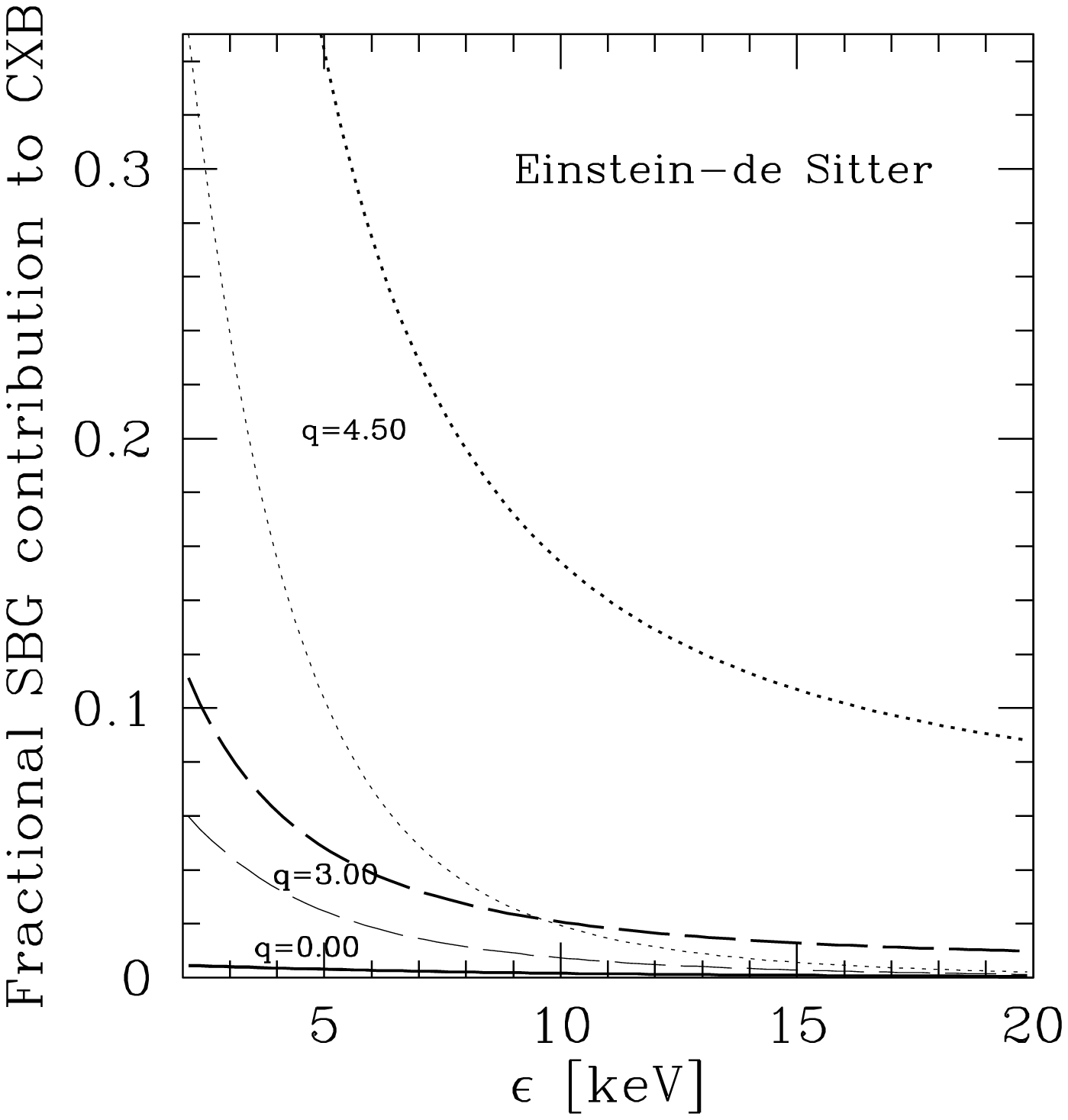}
\vskip 0.2cm
\vspace{6.0cm}
\includegraphics{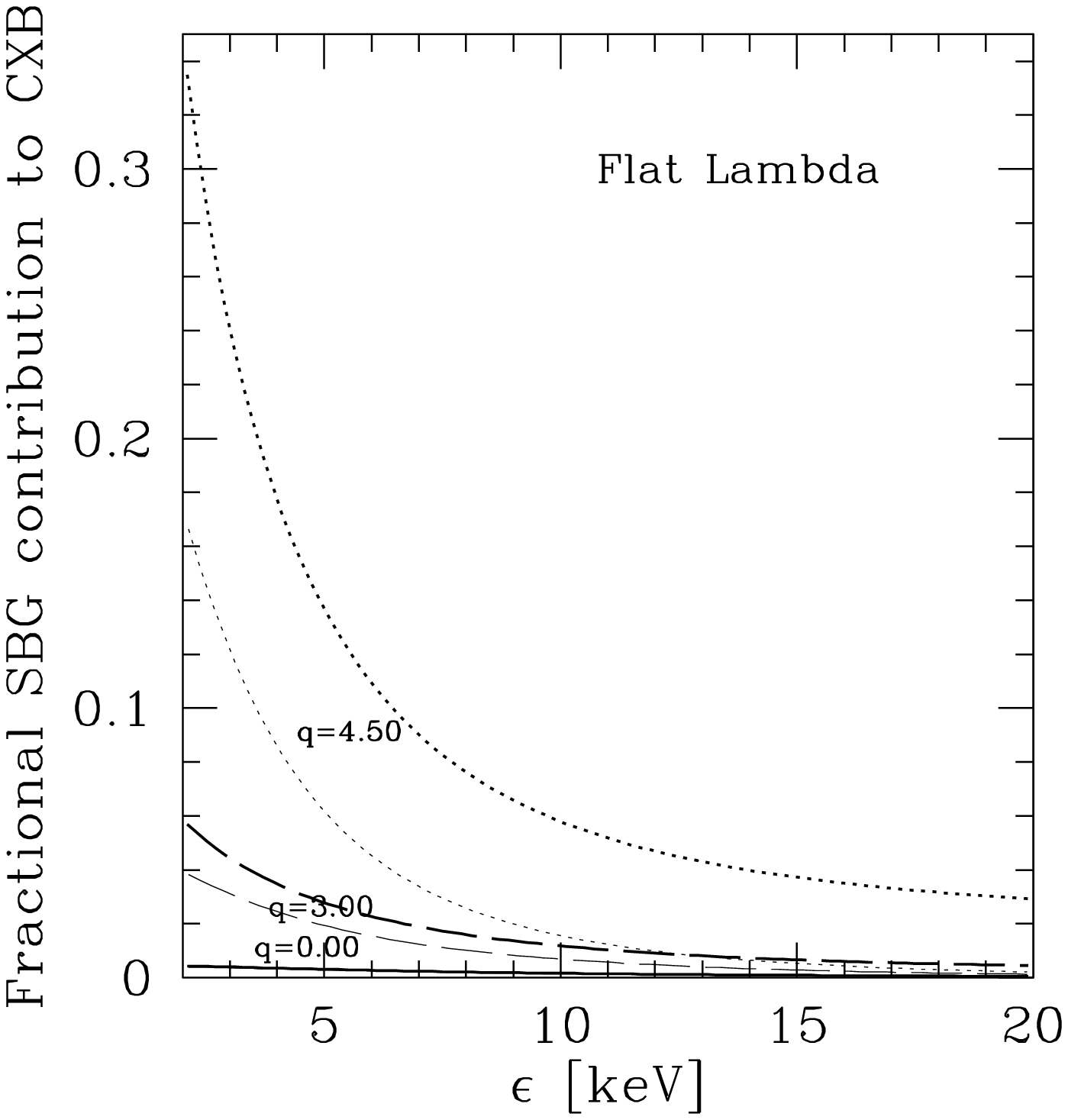}
\caption{
The fraction of the CXB arising from SBGs as a function of energy, for
various strengths of cosmological evolution. Models, parameter values, 
and symbols are as in Fig.3.}
\end{figure}

\section{SBG Emission as a Background}

The surface brightness due to a population of unresolved sources is 
given by (e.g., Boldt 1987)
$$
I ~=~ { 1 \over 4 \pi } \int {\cal L} \, {\rm d}\ell
\eqno(6)
$$
where ${\cal L} \equiv \int L \, \phi(L) \, {\rm d}L$ is the comoving
volume emissivity, and d$\ell$ is the line element. Since d$\ell$ is
related to the look-back time $\t$ and the redshift $z$ by d$\ell =
c/(1+z) {\rm d} \t$, we obtain
$$
I ~=~ { c \over 4 \pi } \int {{\cal L} \, {\rm d} \t \over
1+z }\,.
\eqno(7)
$$
The ensuing spectral density is then given by
$$
{ {\rm d} I \over {\rm d} \e } ~=~ { c \over 4 \pi } \int { {\rm d}
{\cal L} \over {\rm d} \e_0}\, {\rm d} \t \,,
\eqno(8)
$$
where $\e$ is the energy measured in the observer frame and $\e_0 =
\e \,(1+z)$ the energy measured in source rest frame.

The look-back time $\t$ is related to the cosmic time $t$ by $\t \equiv
t_0-t$ (with $t_0$ the present age of the universe), and $t$ is related
to the redshift $z$ by
$$
{{\rm d} t \over {\rm d} z} ~=~ -{1 \over 1+z} ~{1 \over H_0} ~\times ~
$$
$$ ~ \times ~
\sqrt{1 \over
\W_r(1+z)^4 + \W_m(1+z)^3 + \W_\L + (1-\W_0)\,(1+z)^2} \,
$$
where $\W_r$, $\W_m$, $\W_\L$, and $\W_0$ are the current density
parameters for radiation, matter, $\L$, and their sum, respectively.
In the matter dominated era,
$$
{ {\rm d} \t \over {\rm d} z}  = {1 \over H_0}
\times
\left\{
\begin{array}{ll}
~(1+z)^{-2.5} \, & \mbox{\rm EdS} \\
\bigl[ (1+z)  \sqrt{0.3\,(1+z)^3 + 0.7} \,\bigr]^{-1}
 & \mbox{${\rm flat}\,\L$}
\end{array} 
\right.
\eqno(9)
$$

The SBG photon spectrum we have used in our calculations is given in 
eq.(1). Whereas the first two spectral components in eq.(1) have been 
adequately discussed in many previous works, the Compton CMB component 
needs to be further elaborated upon here because of its enhanced role 
in early SBGs. Most of the X-ray emission from a SBG is triggered by 
stellar activity and involves the internal properties of the galaxy 
during a starburst phase of its evolution. As such, the emission is largely 
independent of redshift, if the main characteristics of this phase have 
not changed over cosmic time. The Compton component of the emission has 
evolved significantly, however, due to the much higher energy density 
of the CMB at $z>1$. As we have noted in section 2, in a local SBG 
the FIR energy density is typically much higher than that of the CMB, 
by as much as a factor of $\sim 10$ (Goldshmidt \& Rephaeli 1995), 
so the Compton CMB component begins to dominate already at 
$\sim 1$. This, and the fast evolving SBG density, result in a very 
steep combined spectral and density evolution of the emission from 
SBGs at high X-ray energies where Compton scattering is the main 
emission component in SBGs. Due to the steep $(1+z)^4$ dependence 
of the CMB energy density, $f_{C}^{\rm CMB}(x,z)$ has an explicit 
dependence on $z$ in addition to the implicit dependence through $x$. 

With ${\cal L}(\e_0,z) = {\cal L}(z)\, \phi(\e_0)$, where ${\cal 
L}(z)$ is given in eq.(5) and $\phi(\e_0) \propto \e_0 \,f(\e_0)$ 
is the SBG energy spectrum (source frame) normalized by the 2-10 keV 
flux, we have for the total intensity at an observed energy $\e$ due 
to a population of SBGs undergoing cosmic evolution according to 
eq.(5) out to a maximal redshift $z_m$
$$
I(\e)  = {c\, {\cal L} (0) \over 4\pi H_0}
\cdot
\left\{
\begin{array}{ll}
\int_0^{z_m} \frac{(1+z)^q  \phi[\e(1+z), z]}{(1+z)^{2.5}} {\rm d} z
& \mbox{\rm EdS} \\
    \int_0^{z_m} \frac{(1+z)^q  \phi[\e(1+z), z]}{(1+z)
    \sqrt{0.3 \,(1+z)^3 + 0.7}} {\rm d} z
    & \mbox{${\rm flat}\,\L$}
\end{array}
\right.
\eqno(10)
$$
The integrated intensity from SBGs was calculated at (the source frame)
energies $\e_0$ = 2-120 keV, for values of the evolutionary index $q$ in
the range $0.0 \leq q \leq 4.5$, and adopting $z_m=5$. In Fig.3 we
show the integrated background light arising from an evolving population
of SBGs as a function of the (observed) energy, $\e$. To determine the 
fractional contribution of SBGs to the CXB intensity in the 2-20 keV 
band, for the latter we use the analytic fit:
$$
I_{\CXB}(\e) ~=~
7.877\, \e^{-0.29} \,e^{-{\e / 41.13 }} \,,
\eqno(11)
$$
with $I_{\CXB}$ measured in $\keV/(\cm2 \sec \keV \sr)$ and $\e$ measured 
in keV, given by Gruber et al. (1999) in the 3-60 keV band (based on 
{\sl HEAO-1}/A2+A4 all-sky data: dominated by A2 proportional-counter data for $\e 
< 15$ keV, and by A4 scintillator-experiment data for $\e > 30$ keV) and 
shown by Gendreau et al. (1995) and Chen et al. (1997) to hold down to 1 
keV (from {\sl ASCA} data, based on a composite sample of the sky 
corresponding to less than a square degree and therefore limited by by 
sky surface brightness fluctuations).
	\footnote{Although the spectral shape of the CXB is very well
	established, there is some question about the normalization, 
	especially at $\e \mincir 3$ keV. 
	Under the assumption that Gruber et al.'s (1999) composite 
	multi-experiment fit [which in the 3-50 keV band is very close 
	to the spectrum implied by the {\sl HEAO-1}/A2 data alone 
	(Marshall et al. 1980; Boldt 1987)] is valid down to 1 keV, we 
	note that the {\sl HEAO-1} spectrum is uniformly lower, by 
	$\sim 11\%$, than the PL fit to the {\sl ASCA}/SIS spectrum, 
	$I_{\CXB}(\e) = 8.6\, \e^{-0.33}$ keV/(cm$^2$ s keV sr) 
	(Gendreau et al. 1995). Actually, the discrepancy could be 
	even worse, were the {\sl HEAO-1}/A2 result to be renormalized 
	downwards to account for the fact that it includes relatively 
	bright unresolved sources that would have been resolved by {\sl 
	ASCA}. 
	In addition, the {\sl ROSAT} observed CXB (at $\e > 1$ keV), 
	though limited by surface brightness fluctuations estimated at 
	$\sim 10\%$ (Miyaji et al. 1998), suggests a CXB normalization 
	$\magcir 20\%$ higher than that obtained from the {\sl HEAO-1}/A2 
	result.
	Given this state of affairs, we assume that a smooth extrapolation
	of Gruber et al.'s (1999) {\sl HEAO-1}-based formula down to 2 keV 
	is safe enough for our current purposes.} 
In Fig.4 we plot the SBG fractional contribution to the CXB as a function 
of (observed) energy.

Looking at Fig.3 we see that for energies $\e \magcir 5$ keV the shape of 
the SBG background light gets progressively steeper than that of the CXB, 
for the range of values of $q$ considered here. (The steep rise of the 
Comptonized CMB contribution moderates the decline somewhat.) From Fig.4 
it is seen that for no cosmic evolution ($q=0$) the contribution of SBGs 
to the CXB is $< 1\%$ at all energies; while for an evolution as steep as 
the maximum permitted by IR data ($q \simeq 3$) the contribution is $\sim 
5\%$ at 5 keV, $\sim 2\%$ at 10 keV, $\sim 1\%$ at 15 keV, and vanishing 
at higher energies. This is so for the EdS model; the corresponding results 
are even lower for the flat $\Lambda$ model. 

Also of interest is to determine the $z$ dependence of the superposed SBG 
emission (at a given energy) in order to assess when the contribution is 
maximal. This can be readily found from the integrand of the $z$-integrated 
spectral intensity in eq. (10). Representing the main part of the spectral 
energy flux of a SBG as $\phi(\e_0) \propto \e_0\,f(\e_0) \propto \e_0^{1-\G}$, 
with $\G \sim 2$ in the 1-200 keV band (see Fig.2), we see that in the EdS 
model the integrand in eq. (10) is $\propto (1+z)^{q-1.5-\G}$. Thus, in 
evolving models the relative contribution of SBGs increases with $z$ if $q 
> 3.5$, and decreases if $q < 3.5$. While the critical value of $q$ for which 
this transition occurs is different in the flat-$\Lambda$ model, the behavior 
is qualitatively similar. Note that, as emphasized already, the contribution 
of the Compton component of the flux grows much more steeply due to the fact 
that the CMB energy density is $\propto (1+z)^4$.

\section{Discussion}

The starting point in our treatment and previous similar works 
is the identification of SBGs based on their spectral properties 
at 60$\mu$m, particularly the use of the same estimate for the 
local 60$\mu$m luminosity density, ${\cal L}_{60\mu {\rm m}}$.
However, most previous studies were limited to the consideration 
of only the low-energy ($\leq$2 keV) contribution of SBGs to the CXB. 
Weedman (1987), Griffiths \& Padovani (1990), and Ricker \& Meszaros 
(1993), having based their estimates on {\sl Einstein} data, 
evaluated the SBG contribution at 2 keV. More recently, Moran et al. 
(1999) estimated the SBG contribution to the 5 keV CXB from {\sl 
ASCA} data, based on the assumption that the 5 GHz radio emission 
from galaxies lacking a radio-loud AGN can serve as a gauge of 
star-formation activity, and that the value measured in NGC 3256
for the 5 keV to 5 GHz flux ratio is universal. They used the 5 GHz 
number-count--flux relation, log$N$-log$S_{5\,{\rm GHz}}$, in the 
sub-mJy regime, and estimated that SBGs contribute $\sim 12\% - 28\%$
of the 5 keV CXB.
A less limited assessment of the SBG contribution was implemented 
(Rephaeli et al. 1991, 1995) by considering the SBG emission in the 
2-30 keV band. However, due to lack of observational data, our 
previous works were based on a very preliminary statistical analysis 
of stacked (low-exposure {\sl HEAO-1} A2+A4) data. Here, on the other 
hand, we have estimated the 2-10 keV luminosity density using the 
detailed 60$\mu$m and 2-10 keV luminosities measured for essentially 
the same sample of local SBGs whose X-ray spectra have been adequately 
measured (though not yet at the desired level of precision).

Inclusion of nonthermal emission due to Compton scattering of 
relativistic electrons by the FIR and CMB fields is an important 
new ingredient in the modeling
of the high energy emission in an evolving population of
SBGs. Although the level of this emission is still unknown, our 
explicit accounting for this emission and its expected rapid rise with 
redshift (which makes its relative contribution very
significant at $z>1$) provide further motivation for a more detailed 
description then given here. [The occurrence of low-luminosity AGNs in 
the nuclei of SBGs is a controversial issue (e.g., Veilleux 2001), so 
we chose not to include such a component in the current treatment.]

Our main result that the evolving population of SBGs contributes 
appreciably to the CXB in the 2-10 keV band is quite consistent with 
our previous estimates (Rephaeli et al. 1991, 1995), and a similar 
result by Griffiths \& Padovani (1990). It was argued in the latter 
paper that since the spectrum of X-ray binaries is quite hard and 
comparable in shape to that of the residual CXB in the 2-30 keV range, 
evolving SBGs in the redshift range $0.5 \mincir z \mincir 1$ could 
make an important contribution to the 2-10 keV CXB. In the use of a 
quantitative spectral form and evolutionary function, our treatment 
is similar in spirit to that of Ricker \& Meszaros (1993, hereafter 
RM93). The main improvement with respect to the work of RM93 is the 
use of spectral results from individual SBGs observed with {\sl ASCA} 
and {\sl BeppoSAX} and the inclusion of the two Compton components,
in contrast to their use of essentially an assumed theoretical model. 
[We have included also the observed warm ($\simeq 0.8$ keV) component, 
but its contribution to the emission at $\e > 2$ keV is relatively 
minor.]

In order to check the stability of our results vs. the characteristics of 
the FIR-selected SBG population, we have repeated the calculation using 
the SBG population parameters of Franceschini et al. (2001), who fitted 
the 12$\mu$m number counts invoking the presence of a starbursting 
population whose volume emissivity increases steeply out to $z=0.8$, and 
remains constant at higher $z$ [in a flat $\Lambda$ ($\W_m=0.3$, $\W_\L =
0.7$) cosmological model]. The contribution of Franceschini et al.'s SBG 
population 
	\footnote{ The local 2-10 luminosity density of the 
	starbursting population of Franceschini et al. (2001) 
	is ${\cal L}(0)=3.15 \times 10^{36}$ erg s$^{-1}$ Mpc
	$^{-1}$. This value is deduced from the 12$\mu$m local 
	luminosity function of this population (see their Fig.9), 
	which can be described as $\phi(L) \, {d\, {\rm log}L} ~=~ 
	\phi_\star \, L^{1-a} \, [1+L/(bL_\star)]^{-b} \, {d\, {\rm 
	log}L}$ with $\phi_\star=5.5 \times 10^{-3}$, $L_\star=
	5 \times 10^9 L_\odot$, $a=1.15$, and $b=3.1$, and the 
	average 2-10 keV to 12$\mu$m luminosity ratio of $(1.26 
	\pm 0.35) \times 10^{-3}$, derived for the same sample 
	of local SBGs mentioned in footnote 3 (the 12$\mu$m data 
	are from Shapley et al. 2001). The evolution is ${\cal 
	L}(z) = {\cal L}(0) (1+z)^{7.8}$ for $z \leq z_{\rm break}$ 
	and ${\cal L}(z) = {\cal L}(0) (1+z_{\rm break})^{7.8}$ 
	for $z> z_{\rm break}$, with $z_{\rm break}=0.8$.
	}
turns out to be only marginally higher than that computed for the SBG 
population of Pearson \& Rowan-Robinson (1996) used in our main calculation 
and described by eqs.(2)-(5) (with $q=3$). On the whole, then, the results 
of this variant calculation are qualitatively consistent with our main 
results (see Fig.5, where the 'wind+binaries' spectral model is used). 

Recent results from the {\sl Chandra} deep surveys provide direct 
estimates of the SBG contribution to the CXB. Using the 1 Ms {\sl Chandra}
Deep Field North (Brandt et al. 2001a) and 15$\mu m$ ISOCAM {\sl Hubble} 
Deep Field North (Aussel et al. 1999) surveys, Alexander et al. (2002) 
found a tight correlation between the population of strongly evolving SBGs 
at $z \sim 1$ discovered in faint 15$\mu m$ ISOCAM surveys (e.g., Aussel et 
al. 1999; Elbaz et al. 1999, 2002) -- whose space density is an order of 
magnitude higher than that of local SBGs -- and the apparently normal galaxies 
detected at faint fluxes in X-ray surveys (Giacconi et al. 2001; Hornschemeier 
et al. 2001; Brandt et al. 2001b). The characteristics ($L_{\rm FIR}$, $L_x$, 
2-10 keV spectral slope) of the population of the X-ray detected galaxies, 
whose redshifts are in the range $0.4 \le z \le 1.3$ (based mainly on
Keck data; see Alexander et al. 2002 and references therein), are consistent 
with those expected for SBGs resembling local objects such as M82 and 
NGC 3256. In particular, their stacked average 2-10 keV spectral slope, $\G \sim 2$, 
is steep enough to suggest a low fraction of obscured AGN activity within 
the population. The contribution of this population to the 0.5-8 keV CXB, 
estimated to be $\sim 2\%$ (Alexander et al. 2002), marginally agrees (on 
the low side) with our result for $q=3$ (see Fig.4). A contribution to the 
CXB will also come from the 1 Ms {\sl Chandra} Deep Field North sources 
identified as Lyman-break galaxies at $z \sim 2-4$: these star-forming 
sources have rest-frame 2-8 keV luminosities and X-ray/$B$-band luminosity 
ratios that are comparable to 
those of local SBGs (Brandt et al. 2001c). Based on the above, there is 
no evidence for a substantial change in the $\sim 2-10$ keV luminosity 
and spectral slope of these X-ray-selected, star-forming galaxies. On the 
other hand, their IR (15$\mu m$ ISOCAM) counterparts do show a strong 
density evolution with redshift (e.g., Aussel et al. 1999; Elbaz et al. 
1999, 2002). All this clearly supports our adopted scheme of cosmic 
evolution, which is essentially of the density type (see sect.4). 

Finally, it is commonly thought that AGNs are the prime sources of the CXB 
(e.g.: Comastri et al. 1995; Setti \& Woltjer 1989; Boldt \& Leiter 1981 and 
Leiter \& Boldt 1982), and that additional contributions from other sources 
are typically estimated to be $\mincir 20\%$ in the 2-10 keV range (Giacconi 
et al. 2001; Tozzi et al. 2001; Mushotzky et al. 2000). Based on our results, 
SBGs may possibly be the second most important contributors to the CXB. 
(Clusters of galaxies also contribute a few percent: e.g., Piccinotti et al. 
1982.) The importance of determining the level of the SBG contribution to the 
2-10 keV CXB lies not only in the implications for the evolution of SBGs 
themselves, but also in the ramifications for galaxy evolution in general, 
and as an additional input on the evolution of the AGN phenomenon in 
particular (e.g.: Fabian et al. 1998; Gilli et al. 1999). 

\begin{figure}
\vspace{1.85cm}
\includegraphics{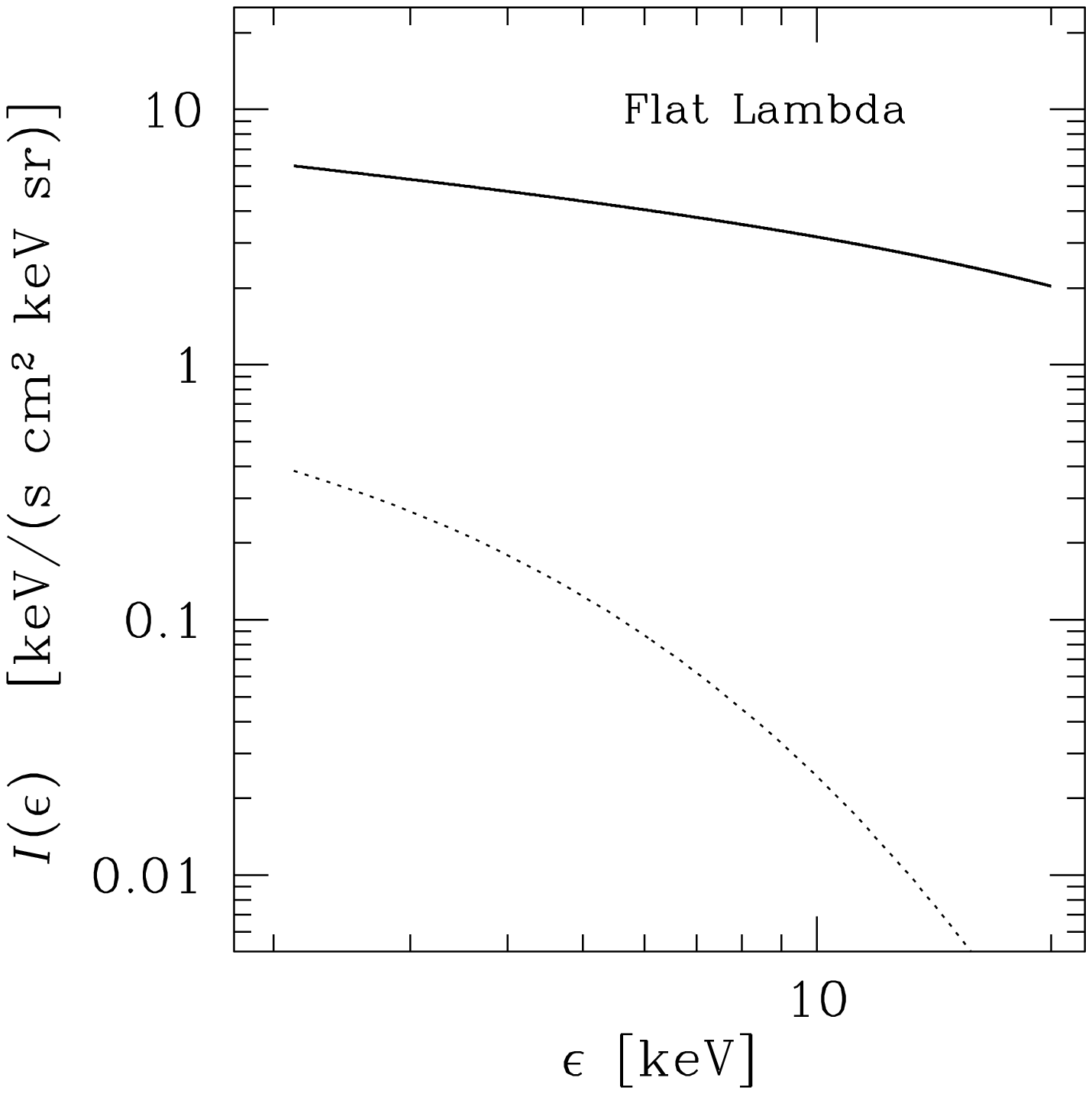}
\hskip 0.2cm
\vspace{1.85cm}
\includegraphics{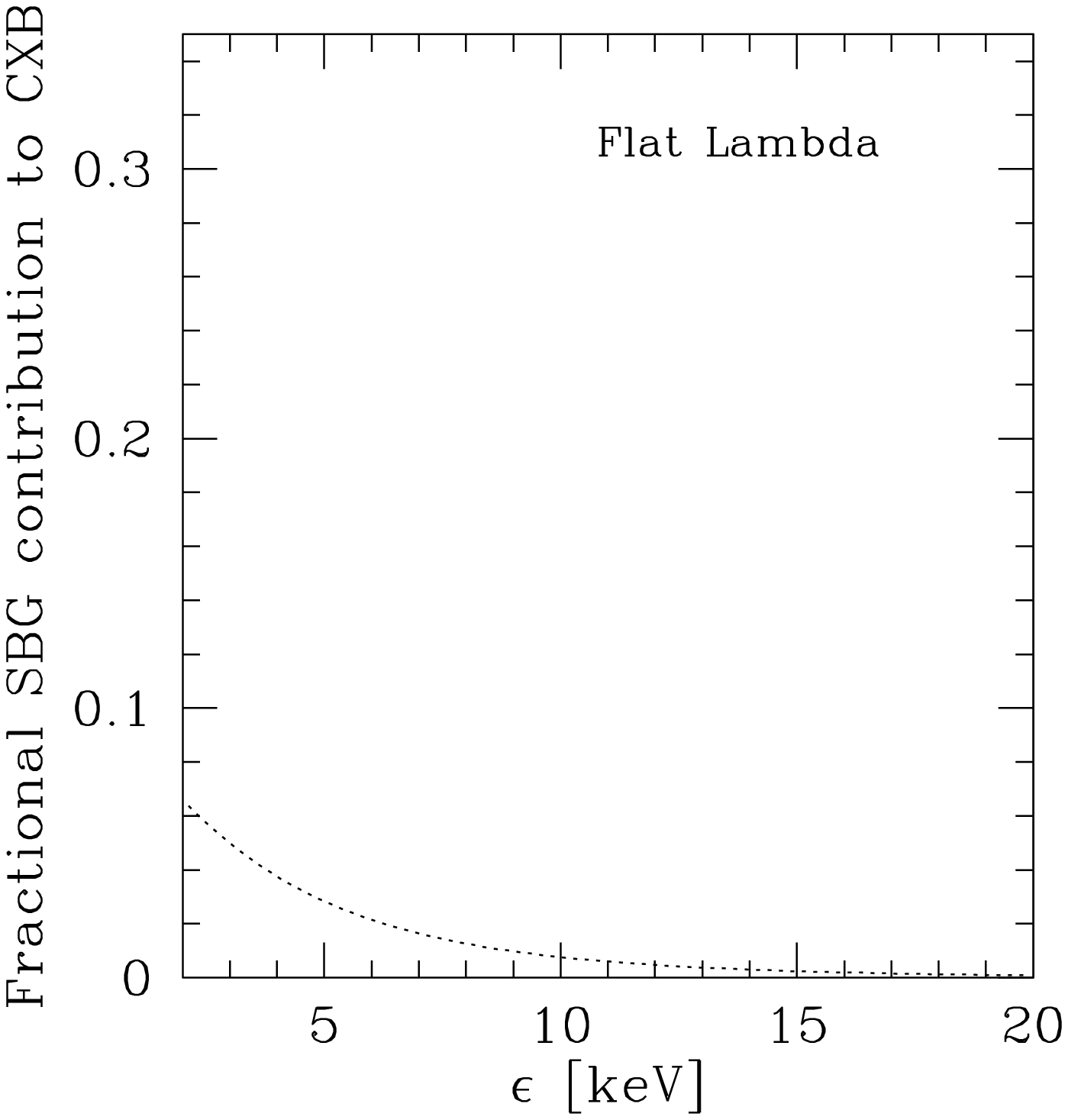}
\caption{
The spectrum of the integrated background light (left), and the 
corresponding contribution to the CXB (right), arising from the 
SBG population of Franceschini et al. (2001), assuming a 'starburst 
only' model for the spectrum. 
See the text for more details.
}
\end{figure}

\begin{acknowledgement} 

We are glad to thank Elihu Boldt for enlightening exchanges on 
the CXB normalization. We also thank an anonymous referee for useful 
suggestions. 
M.P. acknowledges financial support from the Italian Space Agency 
(ASI) through grant No. ASI RS 99 I/R/098/00, as well as the warm 
and stimulating environment of the Center for Astrophysics and Space 
Sciences (CASS) of the University of California at San Diego (UCSD), 
where part of this work was carried out. 
Y.R. acknowledges NASA for supporting his work at UCSD. 
This research has made use of the NASA/IPAC Extragalactic Database 
(NED) which is operated by the Jet Propulsion Laboratory, California 
Institute of Technology, under contract with NASA.

\end{acknowledgement}


\def\ref{\par\noindent\hangindent 20pt} 
 
\noindent 
{\bf References} 
\vglue 0.2truecm

\ref{Alexander, D.M., Aussel, H., Bauer, F.E., Brandt, W.N., Hornschemeier, A.E., 
     Vignali, C., Garmire, G.P., \& Schneider, D.P. 2002, ApJ, 568, L85}
\ref{Aussel, H., Cesarsky, C.J., Ebaz, D., \& Starck, J.L. 1999, A\&A, 342, 313}
\ref{Benn, C.R., Rowan-Robinson, M., McMahon, R.G., Broadhurst, T.J., \& 
     Lawrence, A. 1993, MNRAS, 263, 98}
\ref{Boldt, E. 1987, Phys. Rep., 146, 215}
\ref{Boldt, E., \& Leiter, D. 1981, Nature, 290, 483}
\ref{Bookbinder, J., Cowie, L.L., Krolik, J.H., \& Rees, M. 1980, ApJ, 237, 647} 
\ref{Brandt, W.N., Alexander, D.M., Hornschemeier, A.E., Garmire, G.P.,
    Schneider, D.P., Barger, A.J., Bauer, F.E., Broos, P.S., Cowie, L.L., 
    Townsley, L.K., Burrows, D.N., Chartas, D., Feigelson, E.D., 
    Griffiths, R.E., Nousek, J.A., \& Sargent, W.L.W. 2001a, AJ, 122, 2810}
\ref{Brandt, W.N., Hornschemeier, A.E., Alexander, D.M., Garmire, G.P.,
    Schneider, D.P., Broos, P.S., Townsley, L.K., Bautz, M.W., Feigelson, E.D.,
    \& Griffiths, R.E. 2001b, AJ, 122, 1}
\ref{Brandt, W.N., Hornschemeier, A.E., Schneider, D.P., Alexander, D.M.,
    Bauer, F.E., Garmire, G.P., \& Vignali, C. 2001c, ApJ, 558, L5}
\ref{Cappi, M., Persic, M., Bassani, L., Franceschini, A., Hunt, L.K., Molendi, 
     S., Palazzi, E., Palumbo, G.G.C., Rephaeli, Y., \& Salucci, P. 1999, A\&A, 
     350, 777}
\ref{Chen, L.-W., Fabian, A.C., \& Gendreau, K.C. 1997, MNRAS, 285, 449}
\ref{Christian, D.J, \& Swank, J.H. 1997, ApJS, 109, 177}
\ref{Comastri, A., Setti, G., Zamorani, G., \& Hasinger, G. 1995, A\&A, 296, 1}
\ref{Condon, J.J. 1992, ARAA, 30, 575}
\ref{Dahlem, M., Weaver, K.A., \& Heckman, T.M. 1998, ApJS, 118, 401}
\ref{Danese, L., De Zotti, G., Franceschini, A., \& Toffolatti, L. 1987, ApJ, 
     318, L15}
\ref{David, L.P., Jones, C., \& Forman, W. 1992, ApJ, 338, 82}
\ref{Della Ceca, R., Griffiths, R.E., Heckman, T.M., Lehnert, M.D., \& 
     Weaver, K.A. 1999, ApJ, 514, 772}
\ref{Della Ceca, R., Griffiths, R.E., Heckman, T.M., \& MacKenty, J.W. 
     1996, ApJ, 469, 662}
\ref{Elbaz, D., Cesarsky, C.J., Chanial, P., Aussel, H., Franceschini, A., 
     Fadda, D., \& Chary, R.R. 2002, A\&A, 384, 848}
\ref{Elbaz, D., Cesarsky, C.J., Fadda, D., Aussel, H., Desert, F.X., 
    Franceschini, A., Flores, H., Harwit, M., Puget, J.L., Starck, J.L., 
    Clements, D.L., Danese, L., Koo, D.C., \& Mandolesi, R. 1999, A\&A, 351, L37}
\ref{Ellis, R.S., Colless, M., Broadhurst, T., Heyl, J., \& Glazebrook, K.
     1996, MNRAS, 280, 235}
\ref{Fabbiano, G. 1989, ARAA, ApJ, 330, 672}
\ref{Fabian, A.C., Barcons, X., Almaini, O., \& Iwasawa, K. 1998, MNRAS, 297, L11}
\ref{Franceschini, A., Aussel, H., Cesarsky, C.J., Elbaz, D., \& Fadda, 
     D., 2001, A\&A, 378, 1}
\ref{Franceschini, A., Danese, L., De Zotti, G., \& Xu, C. 1988, MNRAS, 233, 
     175}
\ref{Franceschini, A., Braito, V., Persic, M., Della Ceca, R., Bassani, L.
    Cappi, M., Malaguti, P., Palumbo, G.G.C., Risaliti, G., Salvati, M., 
    \& Severgnini, P. 2002, MNRAS, submitted}
\ref{Fried, J.W., von Kuhlmann, B., Meisenheimer, K., Rix, H.-W., Wolf, C., 
     Hippelein, H.H., K\"ummel, M., Phleps, S., R\"oser, H.J., Thierring, I.,
    \& Maier, C. 2001, A\&A, 367, 788}
\ref{Gendreau, K.C., Mushotzky, R., Fabian, A.C., Holt, S.S., Kii, T., 
     Serlemitsos, P.J., Ogasaka, Y., Tanaka, Y., Bautz, M.W., Fukazawa, Y., 
     Ishisaki, Y., Kohmura, Y., Makishima, K., Tashiro, M., Tsusaka, Y., 
     Kunieda, H., Ricker, G.R., \& Vanderspek, R.K. 1995, PASJ, 47, L5}
\ref{Giacconi, R.,  Rosati, P., Tozzi, P., Nonino, M., Hasinger, G., Norman, C., Bergeron, J., 
     Borgani, S., Gilli, R., Gilmozzi, R., \& Zheng, W. 2001, ApJ, 551, 624}
\ref{Gilli, R., Risaliti, G., \& Salvati, M. 1999, A\&A, 347, 424}
\ref{Goldshmidt, O., \& Rephaeli, Y. 1995, ApJ, 444, 113}
\ref{Griffiths, R., \& Padovani, P. 1990, ApJ, 360, 483}
\ref{Griffiths, R.E., Ptak, A., Feigelson, E.D., Garmire, G., Townsley, L., Brandt, 
     W.N., Sambruna, R., \& Bregman, J.N. 2000, Science, 290, 1325}
\ref{Gruber, D.E., Matteson, J.L., Peterson, L.E., \& Jung, G.V. 1999, 
     ApJ, 520, 124}
\ref{Hornschemeier, A.E., Brandt, W.N., Garmire, G.P., Schneider, D.P.,
     Barger, A.J., Broos, P.S., Cowie, L.L., Townsley, L.K., Bautz, M.W., 
     Burrows, D.N., Chartas, G., Feigelson, E.D., Griffiths, R.E., Lumb, D., 
     Nousek, J.A., Ramsey, L.W., \& Sargent, W.L.W. 2001, ApJ, 554, 742}
\ref{Itoh, N., Sakamoto, T., Kusano, S., Nozawa, S., \& Koyama, Y. 2000, ApJS, 
     128, 125}
\ref{Kim, D.-C., \& Sanders, D.B. 1998a, ApJ, 508, 627}
\ref{Kim, D.-C., \& Sanders, D.B. 1998b, ApJS, 119, 41}
\ref{Leiter, D., \& Boldt, E. 1982, ApJ, 260, 1}
\ref{Lilly, S.J., Le Fevre, O., Hammer, F., \& Crampton, D. 1996, ApJ, 460, L1}
\ref{Lilly, S.J., Tresse, L., Hammer, F., Crampton, D., \& Le Fevre, O. 1995,
     ApJ, 455, 108}
\ref{Lin, H., Yee, H.K.C., Carlberg, R.G., et al. 1999, ApJ, 518, 523}
\ref{Lonsdale, C.J., Hacking, P.B., Conrow, T.P., \& Rowan-Robinson, M. 1990,
     ApJ, 358, 60}
\ref{Lonsdale, C.J., Helou, G., Good, J.C., \& Rice, W. 1985, 'Cataloged 
     Galaxies and Quasars Observed in the \IRAS Survey' (JPL, Pasadena, CA)}
\ref{Madau, P., Ferguson, H.C., Dickinson, M.E., Giavalisco, M., Steidel, C.C., 
    \& Fruchter, A. 1996, MNRAS, 283, 1388}
\ref{Marshall, F.E., Boldt, E.A., Holt, S.S., Miller, R.B., Mushotzky, R.F.,
     Rose, L.A., Rothschild, R.E., \& Serlemitsos, P.J.  1980, ApJ, 235, 4}
\ref{Matsumoto, H., \& Tsuru, T.G. 1999, PASJ, 51, 321}
\ref{Miyaji, T., Ishisaki, Y., Ogasaka, Y., Ueda, Y., Freyberg, M.J., 
    Hasinger, G., \& Tanaka, Y. 1998, A\&A, 334, L13}
\ref{Mizuno, T., Ohbayashi, H., Iyomoto, N., \& Makishima, K. 1998, 
     in 'The Hot Universe', ed. K.Koyama et al., IAU Symp. 188, 284}
\ref{Moran, E.C., Lehnert, M.D., \& Helfand, D.J. 1999, ApJ, 526, 649}
\ref{Morrison, R., \& McCammon, D. 1983, ApJ, 270, 119}
\ref{Mushotzky, R.F., Cowie, L.L., Barger, A.J., \& Arnaud, K.A. 2000, Nature, 404, 459}
\ref{Okada, K., Mitsuda, K., \& Dotani, T. 1997, PASJ, 49, 653}
\ref{Pearson, C., \& Rowan-Robinson, M. 1996, MNRAS, 283, 174}
\ref{Persic, M., Gruber, D.E., \& Rephaeli, Y. 2002, in preparation}
\ref{Persic, M., \& Rephaeli, Y. 2002, A\&A, 382, 843}
\ref{Piccinotti, G., Mushotzky, R.F., Boldt, E.A., Holt, S.S., Marshall, F.E., Serlemitsos, P.J.,
    \& Shafer, R.A. 1982, ApJ, 253, 485}
\ref{Ptak, A., Serlemitsos, P.J., Yaqoob, T., Mushotzky, R., \& Tsuru, T. 1997, 
     AJ, 113, 1286}
\ref{Rephaeli, Y. 1979, ApJ, 227, 364}
\ref{Rephaeli, Y., \& Gruber, D. 2002, A\&A, 389, 752}
\ref{Rephaeli, Y., Gruber, D., \& Persic, M. 1995, A\&A, 300, 91}
\ref{Rephaeli, Y., Gruber, D., Persic, M., \& McDonald, D. 1991, ApJ, 380, L59}
\ref{Ricker, P.M., \& Meszaros, P. 1993, ApJ, 418, 49}
\ref{Roberts, T.P., Schurch, N.J., \& Warwick, R.S. 2001, MNRAS, 324, 737}
\ref{Rothschild, R.E., Baity, W.A., Gruber, D.E., Matteson, J.L., Peterson, L.E.,
    \& Mushotzky, R.F. 1983, ApJ, 269, 423}
\ref{Rowan-Robinson, M., \& Crawford, J. 1989, MNRAS, 238, 523}
\ref{Rowan-Robinson, M., Benn, C.R., Lawrence, A., McMahon, R.G., \&
     Broadhurst, T.J. 1993, MNRAS, 263, 123}
\ref{Saunders, W., Rowan-Robinson, M., Lawrence, A., Efstathiou, G., 
     Kaiser, N., Ellis, R.S., \& Frenk, C.S. 1990, MNRAS, 242, 318}
\ref{Schaaf, R., Pietsch, W., Biermann, P.L., Kronberg, P.P., \& Schmutzler, T.
     1989, ApJ, 336, 722}
\ref{Schmitt, H.R., Kinney, A.L., Calzetti, D., \& Storchi Bergmann, T. 1997, AJ, 114, 592}
\ref{Setti, G., \& Woltjer, L. 1989, A\&A, 224, L21}
\ref{Shapley, A., Fabbiano, G., \& Eskridge, P.B. 2001, ApJS, 137, 139}
\ref{Silva, L., Granato, G.L., Bressan, A., \& Danese, L. 1998, ApJ, 509, 103}
\ref{Stewart, G.C., Fabian, A.C., Terlevich, R.J., \& Hazard, C. 1982, MNRAS, 200, 61P}
\ref{Thompson, R.I., Weymann, R.J., \& Storrie-Lombardi, L.J. 2001, ApJ, 546, 694}
\ref{Tozzi, P., Rosati, P., Nonino, M., Bergeron, J., Borgani, S., Gilli, R., Gilmozzi, R., Hasinger, G.,
     Grogin, N., Keweley, L., Koekemoer, A., Norman, C., Schreier, E., Szokoly, G., Wang, G.X., Zheng, W., 
     Zirm, A., \& Giacconi, R. 2001, ApJ, 562, 42}
\ref{Tucker, W.A. 1975, Radiation Processes in Astrophysics (Cambridge: MIT Press), 169}
\ref{Veilleux, S. 2001, in 'Starburst Galaxies: Near and Far', ed. L.Tacconi \& 
     D.Lutz (Heidelberg: Springer-Verlag), 88}
\ref{Weedman, D.W. 1987, in 'Star Formation in Galaxies', ed. C.J.Lonsdale 
    (NASA CP-2466), 351}
\ref{White, N.E., Swank, J.H., \& Holt, S.S. 1983, ApJ, 270, 711}
\ref{Wilson, G., Cowie, L., Barger, A.J., \& Burke, D.J. 2002, AJ, 124, 1258}
\ref{Zezas, A.L., Georgantopoulos, I., \& Ward, M.J. 1998, MNRAS, 301, 915}

\end{document}